\documentclass[twocolumn,prb,showpacs,floatfix,superscriptaddress]{revtex4}

\usepackage{graphicx}
\usepackage{dcolumn}
\usepackage{float}
\usepackage{amsmath}

\newcommand{\comment}[1]{}

\begin{document}


%
%
\title{\boldmath Marginal Fermi Liquid Analysis of 300~K Reflectance of
Bi$_{2}$Sr$_{2}$CaCu$_{2}$O$_{8+{\delta}}$\unboldmath}

%
%
\author{J. Hwang}
\email{hwangjs@mcmail.mcmaster.ca}
\author{T. Timusk}
\author{A. V. Puchkov}
\affiliation{Department of Physics and Astronomy, McMaster
University, Hamilton, Ontario, Canada L8S 4M1}
\author{N. L. Wang}
\affiliation{Institute of Physics, Chinese Academy of Sciences, P.
O. Box 603, Beijing 100080, People's Republic of China}
\author{G. D. Gu}
\author{C. C. Homes}
\author{J. J. Tu}
\affiliation{Physics Department, Brookhaven National Laboratory,
Upton, New York 11973, USA}
\author{H. Eisaki}
\affiliation{Department of Applied Physics, Stanford University,
Stanford, California 94305}

\date{\today}

%
%
\begin{abstract}
We use 300 K reflectance data to investigate the normal-state
electrodynamics of the high temperature superconductor
Bi$_{2}$Sr$_{2}$CaCu$_{2}$O$_{8+\delta}$ over a wide range of doping
levels.  The data show that at this temperature the free carriers are
coupled to a continuous spectrum of fluctuations.  Assuming the
Marginal Fermi Liquid (MFL) form as a first approximation for the
fluctuation spectrum, the doping-dependent coupling constant $\lambda
(p)$ can be estimated directly from the slope of the reflectance
spectrum. We find that $\lambda (p)$ decreases smoothly with the hole
doping level, from underdoped samples with $ p=0.103$ ($T_c = 67$~K)
where $\lambda (p)= 0.93$ to overdoped samples with $p=0.226$, ($T_c=
60$~K) where $\lambda(p)= 0.53$. An analysis of the intercept and
curvature of the reflectance spectrum shows deviations from the MFL
spectrum symmetrically placed at the optimal doping point
$p=0.16$. The Kubo formula for the conductivity gives a better fit to
the experiments with the MFL spectrum up to 2000 cm$^{-1}$ and with an
additional Drude component or an additional Lorentz component up to
7000 cm$^{-1}$. By comparing three different model fits we conclude
that the MFL channel is necessary for a good fit to the reflectance
data.  Finally, we note that the monotonic variation of the
reflectance slope with doping provides us with an independent measure
of the doping level for the Bi-2212 system.
\end{abstract}


\maketitle


%
%
\section{Introduction}

The complete phase diagram of the high temperature superconducting
(HTSC) cuprates is not yet known. While there is universal
agreement on the presence of an antiferromagnetic phase at low
doping and a superconducting region with a maximum $T_c$ at a
doping level of $p=0.16$ holes per copper site at higher doping,
there is considerable uncertainty about the nature of the {\it
normal state} outside the superconducting phase, particularly in
the overdoped region. It has been clear from the beginning of the
study of HTSC materials that their normal state is highly
anomalous and not at all like the Fermi-liquid state of the
conventional superconductors. Three features stand out. First, at
high temperature, the resistivity and the optical reflectance are
dominated by scattering processes with a linear variation of the
scattering rate with frequency and temperature where $1/\tau
\simeq \max(\omega,\pi T)$.\cite{anderson88,anderson89,varma89}
Experimentally, these processes manifest themselves as a dc
resistivity that varies linearly with temperature with a zero
intercept on the temperature axis\cite{gurvitch87} and an infrared
reflectance that varies linearly with frequency over a very wide
range of frequencies, well into to the
midinfrared.\cite{schlesinger90,rotter91} This singular behavior
of the scattering rate has been termed Marginal Fermi Liquid (MFL)
behavior from the original postulate of Varma {\it et
al.}~\cite{varma89} where it was assumed that the quasiparticle
weight vanished logarithmically at the Fermi surface, and the
Green's function of the carriers was entirely incoherent.

The second anomalous feature of the normal state is the presence
of a pseudogap, a depression in the density of states below a
temperature $T^*$ that decreases as the doping level $p$ is
increased.\cite{pseudogap_review} The pseudogap region on the
phase diagram is not well defined and the exact value of $T^*$
depends on the experimental probe used. Photoemission and
tunneling measurements of Bi$_2$Sr$_2$CaCu$_2$O$_{8+\delta}$
(Bi-2212) point to a $T^* > 300$ K for underdoped
samples,\cite{krasnov02} and below 300 K for optimally-doped
samples.\cite{krasnov02,renner98,krasnov00,suzuki00} In optical
spectroscopy the pseudogap is seen as a depression of interplane
({\it c}-axis) conductivity, which in underdoped YBCO with
$T_c=63$ K, $T^{*}$ occurs below 300 K
(Ref.~\onlinecite{homes93a}). The NMR shift also begins to drop
below its high temperature value near 300 K
(Ref.~\onlinecite{takigawa91}). In Bi-2212 angle-resolved
photoemission (ARPES) measurements show that the pseudogap
develops below 250 K in slightly underdoped Bi-2212
(Ref.~\onlinecite{damascelli03}). Probably most accurate and
recent results will be a c-axis transport measurement of
Bi-2212~\cite{shibachi01}. In the literature $T^*$ is around 300 K
at $p=0.16$ with rather large error bars in the underdoped phases.
In all, it appears that a temperature of 300 K is above $T^*$ for
optimally and overdoped states and below $T^*$ for underdoped
states in the Bi-2212 system.

The third normal-state anomaly is the development of a sharp
scattering resonance in the {\it ab}-plane transport properties
 at low
temperature.  It manifests itself as a sharp increase in the {\it
ab}-plane scattering rate at 500 cm$_{-1}$ that takes place below
150~K in underdoped materials and near $T_c$ in optimally doped
ones.\cite{puchkov96d,timusk03} It is also seen as a sharp kink
in the ARPES dispersion curves.\cite{damascelli03}

The aim of this work is to investigate, in a systematic way, the
transport properties in Bi-2212 for a wide range of doping levels
through accurate reflectance spectroscopy at 300~K. We restrict
our work to the high-temperature strange-metal region where the
spectra are dominated by the MFL fluctuations avoiding the low
temperatures near $T_c$ and where these fluctuations are gapped by
the appearance of the resonant mode. By confining our measurements
to high temperatures, we also steer away from the problem of the
phase diagram in the overdoped state. We have chosen Bi-2212 for
our measurements for several reasons.  First, this material has
been widely investigated by the angle resolved photoemission and
tunneling spectroscopy. Secondly, large single crystals can be
grown and annealed to yield samples spanning the overdoped as well
as the underdoped regions of the phase diagram.

Previous work on the optical properties of Bi-2212 has
concentrated on the optimally doped
materials.\cite{romero91,elazrak94,puchkov96d,baraduc96,
quijada99,tu02} The focus has been on analyzing the optical
conductivities or the scattering rates. Kubo-MFL analysis (see
Sec.~II-D for a detail description) has been used by Littlewood
{\it et al.},\cite{MFL2} Baraduc {\it et al.},\cite{baraduc96} and
Abrahams.\cite{abrahams96}  To improve the overall accuracy, we
fit the reflectance curves to models, avoiding the additional
Kramers-Kronig (K-K) analysis required to calculate the
conductivity and the scattering rates (see Sec. II-D). We also
produce the optical conductivities by using K-K analysis fit them
and compare the results with those of reflectance fits. Since the
reflectance is related to the scattering rate in a direct way,
models where the scattering rate has a simple form, such as the
MFL hypothesis, allow us to estimate the coupling constant
$\lambda(p)$ directly from the reflectance data.

The present paper is organized as follows. First, we introduce a
simple way to extract approximate MFL parameters directly from the
reflectance data. Secondly, we report indications of deviations
from the simple MFL form from the study of accurate 300 K
(Bi-2212) reflectance data in a wide range of doping levels. The
data presented here focus on the overdoped region where we present
new data for several highly overdoped samples ($T_{c}$= 82, 73, 65
and 60~K). We also included data from earlier
measurements,\cite{puchkov96c, puchkov96d} as well as work from Tu
{\it et al.}\cite{tu02} As mentioned above, we focus on 300~K
reflectance data for our analysis, first because we want to avoid
the region where the sharp scattering resonance distorts the
spectra and secondly because with our measurement technique the
data at room temperature are the most accurate.

%
%
\section{Experimental Data and Analysis}
%
%
\subsection{Experimental Techniques}

We used Fourier-transform infrared (FTIR) spectroscopy to obtain
the reflectance data of floating zone grown single twinned
crystals of Bi-2212. To obtain overdoped samples the crystals were
annealed in 3~kbar liquid oxygen in sealed
containers.\cite{mihali93} A polished stainless steel mirror was
used as an intermediate reference to correct for instrumental
drifts with time and temperature. An in-situ evaporated gold film
on the sample was the final reflectance reference.\cite{homes93b}
The reflectance of the gold films was in turn calibrated with a
polished stainless steel sample where we relied on Drude theory
and the dc resistivity as the ultimate reference. An advantage of
this technique is that it corrects for geometrical effects of an
irregular surface. The in-situ gold evaporation technique gives
accurate, better than $\pm 0.5$~\%, room temperature data.

%
%
\subsection{Trends in Reflectance}

\begin{figure}[t!]
  \vspace*{-0.4cm}%
  \centerline{\includegraphics[width=3.5in]{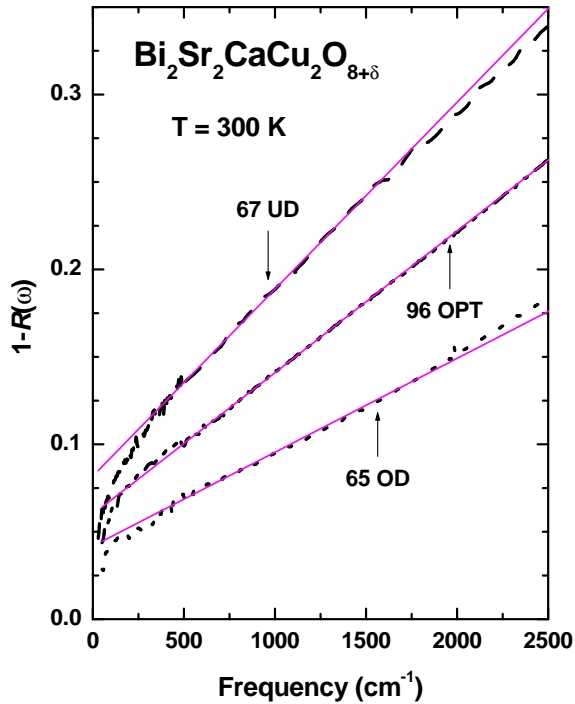}}%
  \vspace*{-1.0cm}%
\caption{Absorption ($1-R$) data at 300~K for
Bi$_2$Sr$_2$CaCu$_2$O$_{8+\delta}$ at three representative doping
levels: one underdoped ($T_{c}=67$~K) sample, one optimally-doped
($T_{c}=91$~K) sample, and one overdoped ($T_{c}=65$~K) sample and
their linear fits, shown as solid lines. We note that both the
slope and the intercept of $1-R$ decrease as the doping level increases.}%
\label{Reflec}
\end{figure}

\begin{figure}[t!]
  \vspace*{-0.4cm}%
  \centerline{\includegraphics[width=3.5in]{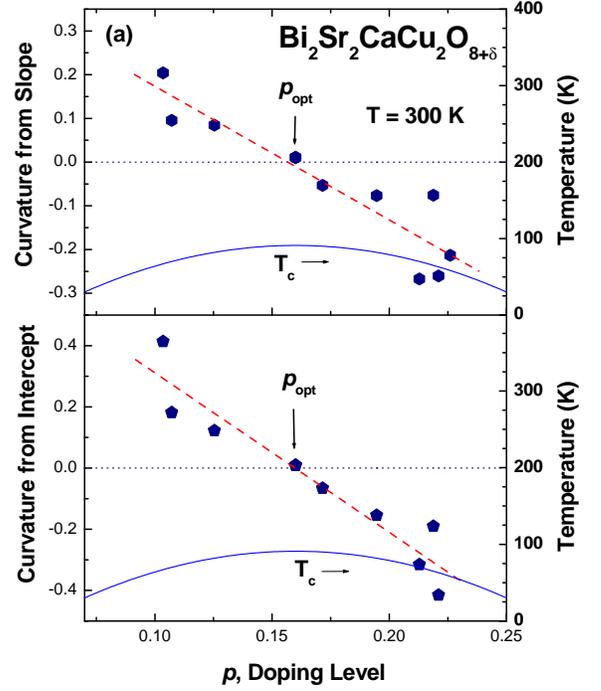}}%
  \vspace*{-1.0cm}%
\caption{Two doping dependent curvature of $1-R$ of
Bi$_2$Sr$_2$CaCu$_2$O$_{8+\delta}$ at 300~K. The curvatures
of the $1-R$ curves go through zero at optimal doping where $p=0.16$
(see Sec.~II-B for a detailed description).}%
\label{curvature}
\end{figure}

Fig.~\ref{Reflec} shows the overall trends of the absorption $(A
\equiv 1-R)$ with doping.  Three representative samples are shown:
an underdoped ($T_{c}=67$~K) sample, an optimally doped
($T_{c}=91$~K) one, and finally an overdoped ($T_{c}=65$~K)
sample. We note that the absorption varies linearly with frequency
at high frequency for all doping levels and that both the slopes
of the curves and their intercepts with the absorption axis
decrease with increasing doping. We also note that the optimally
doped reflectance shows best linearity up to high frequency (see
Fig.~\ref{curvature}) but still we can use this universal linear
trend in the relaxation region as {\it a first approximation}. The
actual reflectance should have an additional second order doping
dependent term, which we discuss with curvature analysis of the
reflectance in the end of this subsection. Within the Drude model
it is easy to show that in the relaxation regime ($1/{\tau}\ll
\omega \ll {\omega}_{p}$), where the imaginary part of the
refractive index is much larger than the real part, the absorbance
$A$ is given by $A=2/{\omega}_{p}\tau$
(Ref.~\onlinecite{timusk89}), where $1/\tau$ is the scattering
rate, $\omega$ the frequency, and ${\omega}_{p}$ the plasma
frequency. Thus the linear variation of absorption suggests a
linear variation of scattering rate $1/\tau$ with frequency. We
also show least-squares fits to straight lines in the frequency
range from 500 to 1750~cm$^{-1}$ in the figure. We will call this
analysis ``$1-R$ slope analysis''.

At low frequency, below the relaxation region where $\omega \ll
1/{\tau}$, the absorption drops below the fitted lines.  This is
the Hagen-Rubens region where the reflectance varies as
$\sqrt{\omega}$. We note that the Hagen-Rubens frequency range
gets smaller as the doping level increases, which confirms the
notion that the scattering rate $1/\tau$ decreases with doping,
since within the Drude model the changeover from Hagen-Rubens
region to the relaxation region occurs at $\omega\tau = 1$.
%
%
%
%
\begin{figure}[t!]
  \vspace*{-0.4cm}%
  \centerline{\includegraphics[width=3.5in]{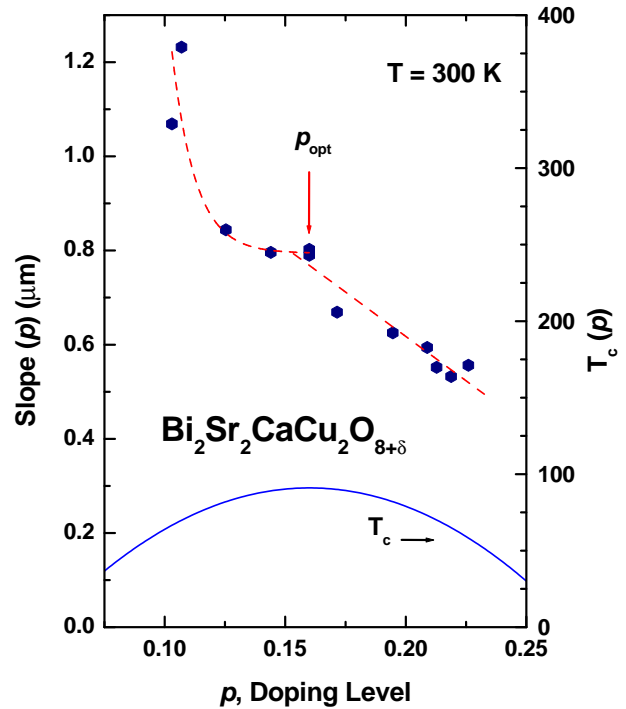}}%
  \vspace*{-1.0cm}%
\caption{Doping dependent slope of the absorption curves shown in
Fig.~\ref{Reflec}, $S(p)$, of
Bi$_{2}$Sr$_{2}$CaCu$_{2}$O$_{8+{\delta}}$ a 300 K. We observe two
different trends: a rapid drop of slope with doping in the
underdoped region and a slower linear variation in the overdoped
region with a crossover point at $p=p_{opt}=0.16$. The $T_{c}$ vs.
$p$ curve shows where our systems are in the phase diagram.}%
\label{slope}
\end{figure}

We determined the doping level $p$ from $T_{c}$ using the
parabolic expression of Presland {\it et al.}\cite{presland91} The
expression is $p(T_{c}) = 0.16
\mp[{1}/{82.6}\,(1-{T_{c}}/{T_{c}^{max}})]^{1/2}$, where
$T_{c}^{max}$ is the maximum $T_{c}$ or $T_{c}(p=0.16)$ and $\mp$
means that we use $-$ for underdoped samples and $+$ for overdoped
samples. The determination of $T_{c}^{max}$ is a delicate
problem\cite{eisaki02} and, in the absence of a better method, we
use the generally accepted value of 91~K as the $T_{c}^{max}$ for
Bi-2212. We should mention here that we have one optimally-doped
sample which is doped with additional small amount of Y to yield a
relatively well ordered system and shows a surprisingly high $T_{c}=96$
 K~\cite{eisaki02}. The disadvantage of the Presland method is
that it does not uniquely determine the doping level of the sample
since there are two independent $p$ values for each value of
$T_{c}$. However, as Fig.~\ref{Reflec} shows, the reflectance
slopes vary monotonically with doping and provide an alternate
method of determining the doping level that does not suffer from
this ambiguity.

As a first step, we fit the $1-R$ data to a straight line in the
frequency range between 500 and 1750~cm$^{-1}$ to get the slopes
and intercepts at various doping levels. Fig.~\ref{slope} shows
the doping dependence of the slope, $S(p)$ obtained from the $1-R$
curves this way. We note the smooth variation of the slope as the
doping level changes. Two different trends can be discerned: a
rapid decrease in the underdoped region and a slower decreasing
trend in the overdoped region.

%
%
\begin{figure}[t!]
  \vspace*{-0.4cm}%
  \centerline{\includegraphics[width=3.5in]{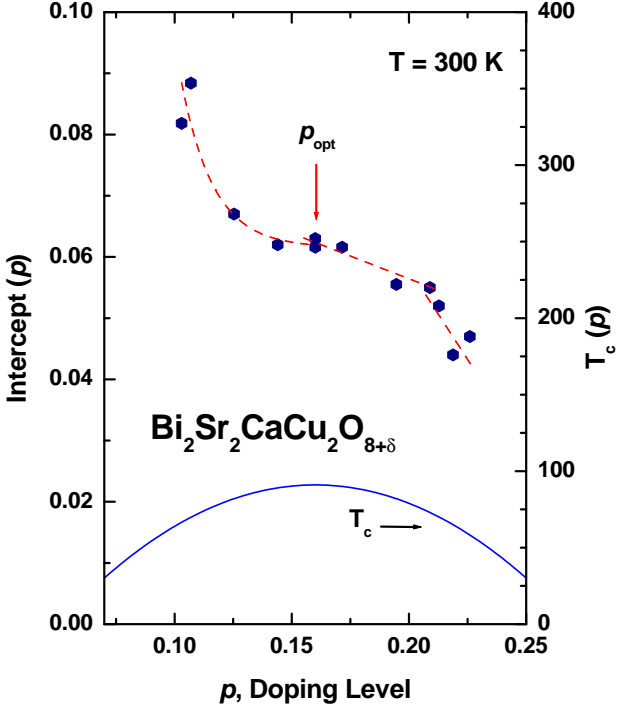}}%
  \vspace*{-1.0cm}%
\caption{Doping dependent intercept, $I(p)$, of
Bi$_2$Sr$_2$CaCu$_2$O$_{8+\delta}$ at 300~K from the fits (see
Table~\ref{Bi2212}). We observe three different trends: an abrupt
decay in the underdoped region, a slow drop between
$p=p_{opt}=0.16$ and $p=p_{c}=0.21$, and an abrupt drop above
$p=p_{c}$ as the doping level increases. We observe two crossover
points at $p=p_{opt}$ and $p=p_{c}$. We also show the $T_{c}$ vs.
$p$ curve.}%
\label{intercept}
\end{figure}

%
%

Fig.~\ref{intercept} shows a doping dependent intercept, $I(p)$ of
$1-R$ on the $\omega=0$ axis. The intercept is related to the
scattering rate at $\omega=0$ and to the dc resistivity. The
intercept also changes smoothly as the doping level changes. The
intercepts are consistent with the slopes within the MFL
hypothesis, except at the high doping range, where we observe a
hint of a crossover around $p=0.21$. There are three regions: a
sharp decaying trend in the underdoped region similar to what we
observed in the slope, a slower decrease  in the optimally doped
region similar to what we observed in the variation of the slope.
At $p \geq 0.21$ we see a crossover where is the intercept drops
more abruptly, i.e., the scattering rate drops quickly.

We found another interesting doping dependent trend: a slight
deviation from a straight line reflectance in the relaxation region
giving rise to a curvature of the reflectance data. We analyze the
curvatures of $1-R$ at various doping levels in the following way: We
take two frequency segments, $500-1250$~cm$^{-1}$ and
$1500-2250$~cm$^{-1}$ in the relaxation region. Then we fit each
segment to a straight line.  Next we get slopes and intercepts of the
two fitted straight lines. We label these $S_{L}$ and $I_{L}$ (of the
lower frequency section) and $S_{H}$ and $I_{H}$ (of the higher
frequency section). We estimate two curvatures independently for each
$1-R$ curve: $C_S=({S_{L}-S_{H}})/{S_{L}}$ from the slopes and
$C_I=-({I_{L}-I_{H}})/{I_{L}}$ from the intercepts.
Fig.~\ref{curvature} shows the two doping dependent curvatures. A
positive (negative) curvature stands for an upward convex (concave)
curve. As the figure shows, the curvature of $1-R$ is positive for the
underdoped samples, negative for the overdoped samples and goes
through zero exactly at optimal doping where $1-R$ is a straight line.

%
%
\subsection{Analysis of \boldmath $1-R$ \unboldmath}
The trends which we observe in Fig.~\ref{Reflec}, the linear
variation of the scattering rate with frequency and the overall
decrease in the slopes and intercepts with doping, can be related
to the parameters of the MFL theory where the  scattering rate,
$1/{\tau}({\omega},p) \approx$ ${\lambda}(p)[{\omega}+{\pi}T ]$,
where $\lambda$ is the coupling constant, $p$ is a doping level,
and $T$ is the temperature.\cite{MFL2}

For thick enough superconducting samples the absorbance ($A$) is
$1-R$. In the relaxation regime we have the following:

\begin{equation}
  A({\omega},p)=\frac{2}{\omega_{p}(p)\tau(\omega, p)}
  =\frac{2\lambda_{S}(p)}{{\omega}_{p}(p)}{\omega}+\frac{2\ \lambda_{I}(p)\pi
  T}{{\omega}_{p}(p)}.%
\label{e03}
\end{equation}

The above equation is linear in frequency for a fixed doping
level. We get the doping dependent slope, $S(p)$, and intercept,
$I(p)$, from a least square fit of our $1-R$ data to a straight
line.  According to Eq. (1) the slopes and intercepts are related
to each other:  $\pi T S(p)=I(p)$ and therefore we would obtain
the same value of the doping dependent coupling constant
$\lambda(p)$ from either the slopes or the intercepts. To account
for deviations from Eq. (1) in our approximate analysis we allow
the two constants to be different. We call the coupling constants
from the slope $\lambda_{S}(p)$ and from the intercept
$\lambda_{I}(p)$. Within the MFL hypothesis
$\lambda_{S}(p)=\lambda_{I}(p)$. In terms of the measured
quantities $S(p)$ and $I(p)$ they are given by:

\begin{equation}
  \lambda_{S}(p)=\frac{1}{2}{\omega}_{p}(p)\,S(p),
  \:\:\:
  \:\:\:
  \lambda_{I}(p)=\frac{I(p){\omega_{p}}(p)}{2\pi T}.%
\label{e04}
\end{equation}

\begin{figure}[t!]
  \vspace*{-0.4cm}%
  \centerline{\includegraphics[width=3.5in]{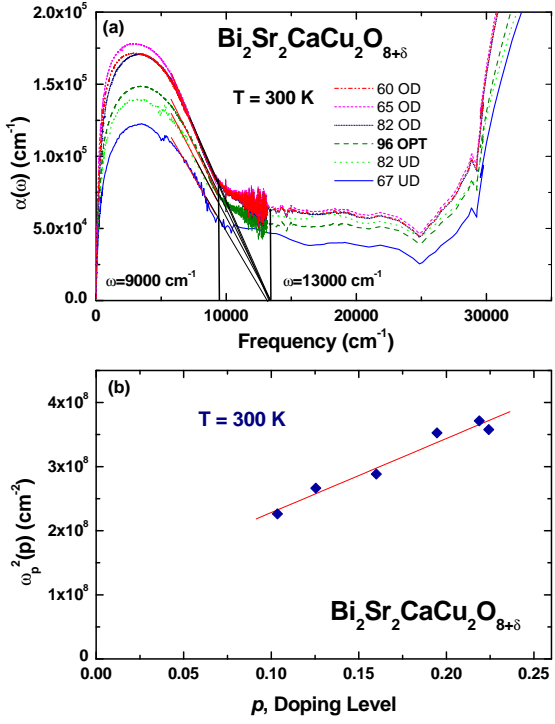}}%
  \vspace*{-1.0cm}%
\caption{(a) Absorption coefficient $\alpha$ at 300~K for
Bi$_{2}$Sr$_{2}$CaCu$_{2}$O$_{8+{\delta}}$ at six representative
doping levels: two underdoped ($T_{c}=67$ and 82 K), one optimally
doped ($T_{c}=96$~K), and two overdoped ($T_{c}=82$, 65, and 60~K)
samples. (b) Squares of the doping-dependent plasma frequencies
$\omega_p^2$ of Bi-2212 at 300~K,
calculated by using Eq.~(\ref{plasm}). The plasma frequency increases monotonically with doping.}%
\label{Absorp}
\end{figure}

To get the absolute value of coupling constants $\omega_{p}(p)$
must be known. We estimate $\omega_{p}(p)$ from the spectral
weight of the conductivity below the interband-transition region.
To determine the frequency of the onset of interband transitions,
we note that the absorption coefficients, $\alpha(\omega) =
4\pi\sigma(\omega)/[n(\omega)c]$, begin to deviate from the low
frequency form at 9000~cm$^{-1}$ and extrapolate to zero at the
same frequency (13000~cm$^{-1}$) for all the samples, independent
of doping, as shown in the upper panel of Fig.~\ref{Absorp}.  This
fact suggests the following method of estimating the spectral
weight up to the interband transition:
\begin{equation}
  \omega_{p}^{2} = \frac{120}{\pi} \left[ \int^{9000}_{0}\sigma_{1}(\omega)d\omega+
  \frac{1}{2}\int^{13000}_{9000}\sigma_{1}(\omega)d\omega \right] .%
  \label{plasm}
\end{equation}

In the lower panel of Fig.~\ref{Absorp} we show the resulting
plasma frequencies of Bi-2212 at 300~K obtained this way. We see
that the plasma frequency increases monotonically as the doping
level increases.\cite{kenziora_comment}

%
%

%
%
\begin{figure}[t!]
  \vspace*{-0.4cm}%
  \centerline{\includegraphics[width=3.5in]{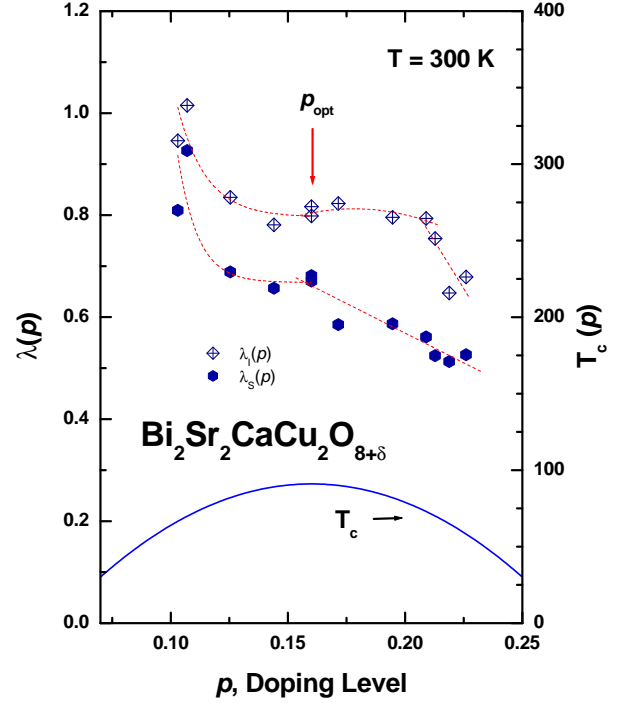}}%
  \vspace*{-1.0cm}%
\caption{Doping-dependent coupling constants, $\lambda_{S}(p)$ and
$\lambda_{I}(p)$, of Bi$_2$Sr$_2$CaCu$_2$O$_{8+\delta}$ at 300~K.}
\label{coupl}
\end{figure}

Fig.~\ref{coupl} shows the doping-dependent coupling constants,
$\lambda_{S}(p)$ and $\lambda_{I}(p)$ obtained using
Eq.~(\ref{e04}). Since the doping-dependent plasma frequency
increases smoothly with the doping level we see the same trends in
$\lambda_{S}(p)$ and $\lambda_{I}(p)$ that we observed in the
doping-dependent slope and intercept.  We note that the coupling
constant $\lambda_{I}(p)$ is larger than $\lambda_{S}(p)$. This is
not surprising since in our simple analysis we have neglected the
frequency dependence of the plasma frequency required by
causality. We correct this in the more complete analysis in the
next section. The simple analysis shows an overall trend of a
decrease of the dimensionless coupling constants as the doping
level increases, which is consistent with the angle-resolved
photoemission results.\cite{johnson01}

%
%
\subsection{Kubo formula in MFL and Kubo-MFL fit}

\subsubsection{Reflectance and Kubo-MFL fit}
%
%
\begin{figure}[t!]
  \vspace*{-0.4cm}%
  \centerline{\includegraphics[width=3.5in]{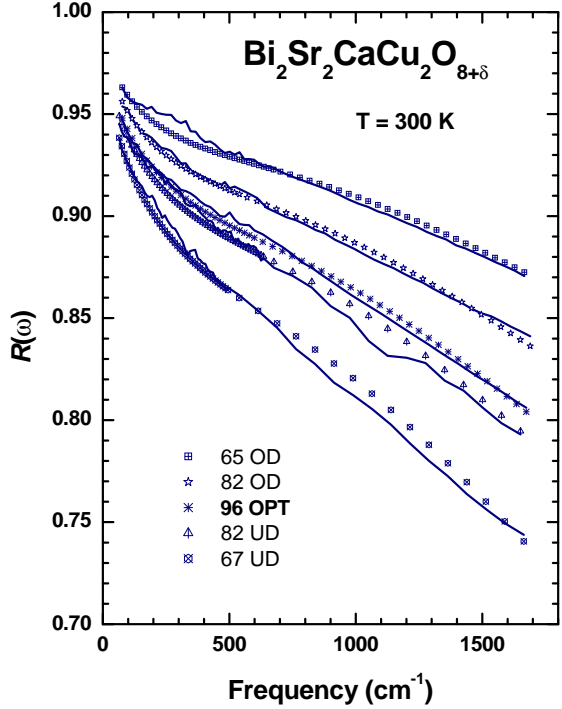}}%
  \vspace*{-1.0cm}%
\caption{Kubo-MFL fits at 300~K for
Bi$_{2}$Sr$_{2}$CaCu$_{2}$O$_{8+{\delta}}$ at five representative
doping levels: two underdoped ($T_{c}=67$ and 82~K), one optimally
doped ($T_{c}=96$~K), and two overdoped ($T_{c}=82$ and 65~K)
samples. Symbols are fits and solid lines are corresponding data.
The fit parameters are shown in Table~\ref{tKubo}.}%
\label{Kubo}
\end{figure}

%
%
\begin{table}
\caption{Kubo-MFL fit parameters of five representative
Bi$_2$Sr$_2$CaCu$_2$O$_{8+\delta}$ samples: two underdoped
($T_{c}=67$ and 82~K), one optimally doped ($T_{c}$=96~K), and two
overdoped ($T_{c}=82$ and 65~K) samples. The results are shown in
Fig.~\ref{Kubo}.}
%
%
%
\begin{ruledtabular}
\begin{tabular}{ccccc}
%
%
  $T_{c}$\ (K) & $\omega_{c}$\ (cm$^{-1}$) & $\lambda_{K}$
  & $\omega_{p}$\ (cm$^{-1}$) & $\epsilon_{\infty}$ \\
  \hline
  67 & 2100 & 0.899 & 15040 & 3.14 \\
  82 & 2100 & 0.720 & 16320 & 3.21 \\
  96 & 2100 & 0.693 & 16980 & 3.52 \\
  82 & 2100 & 0.606 & 18789 & 4.16 \\
  65 & 2100 & 0.457 & 19270 & 4.25 \\
\end{tabular}
\end{ruledtabular}
\label{tKubo}
\end{table}

Here we introduce a more accurate method of analysis.  One can
derive the complex optical conductivity from the Kubo formula and
the MFL hypothesis as follows.\cite{MFL2,abrahams96}
%
%
\begin{eqnarray}
  \tilde\sigma(\omega) \! & = & \! -i\frac{\omega_{p}^{2}}{4 \pi}\frac{1}{2\omega}
    \!\int^{\infty}_{-\infty} \!\!\!\! dy[\tanh \frac{\beta \
    (y+\omega)}{2}\!-\!\tanh \frac{\beta \ y}{2}] \nonumber \\
  & \  & \times [\frac{1}{\Sigma^{R}(y+\omega)-\Sigma^{A}(y)-\omega}],
\label{Kubo1}
\end{eqnarray}
where
\begin{eqnarray}
\Sigma^{R}(\omega) & = &\lambda\omega \ln
\frac{x}{\omega_{c}}-i\frac{\pi
\lambda}{2}x, \nonumber \label{Kubo2} \\
\Sigma^{A}(\omega) & = &\Sigma^{R}(\omega)^{*}, \\
 x & = &\max(|\omega|, \pi T). \nonumber
\end{eqnarray}
Here $\Sigma^{R}(\omega)$ and $\Sigma^{A}(\omega)$ are the
retarded and advanced single-particle self-energies, respectively,
$\omega_{c}$ is the cutoff frequency and $\beta \equiv 1/k_{B}T$.

\begin{eqnarray}
 \tilde\epsilon(\omega)& = & i \frac{4\pi}{\omega}\tilde\sigma(\omega) + \epsilon_{\infty} \nonumber \\
 R(\omega) & = & |\frac{\sqrt{\tilde\epsilon(\omega)}-1}{\sqrt{\tilde\epsilon(\omega)}+1}| ^{2}. \label{Kubo3}
\end{eqnarray}
where $\tilde\epsilon(\omega)$ is the complex dielectric function,
$R(\omega)$ is the reflectance and $\epsilon_{\infty}$ is the
background dielectric constant. We will refer to a fit using
Eqs.~(\ref{Kubo1}), ~(\ref{Kubo2}) and (\ref{Kubo3}) as a
``Kubo-MFL fit''.  We fixed the plasma frequencies,
$\epsilon_{\infty}$ and the cut-off frequency, leaving only
$\lambda$ as a free parameter for each doping level.

Fig.~\ref{Kubo} shows five representative data sets [$T_{c}=67$~K
(UD), $T_{c}=82$~K (UD), $T_{c}=96$~K (OPT), $T_{c}=82$~K (OD),
and $T_{c}=65$~K (OD)] and their Kubo-MFL fits between 60 and
1700~cm$^{-1}$ and Table~\ref{tKubo} shows parameters for the
fits.  One interesting feature of the fits is that the fitted
curves have slight curvature while the reflectance data are
straighter. The curvature of the fits above 500~cm$^{-1}$ is
convex upward (see Sec.~II-B) supporting the argument above. The
pseudogap induces a concave upward curvature on the reflectance.

%
%
\begin{figure}[t!]
  \vspace*{-0.4cm}%
  \centerline{\includegraphics[width=3.5in]{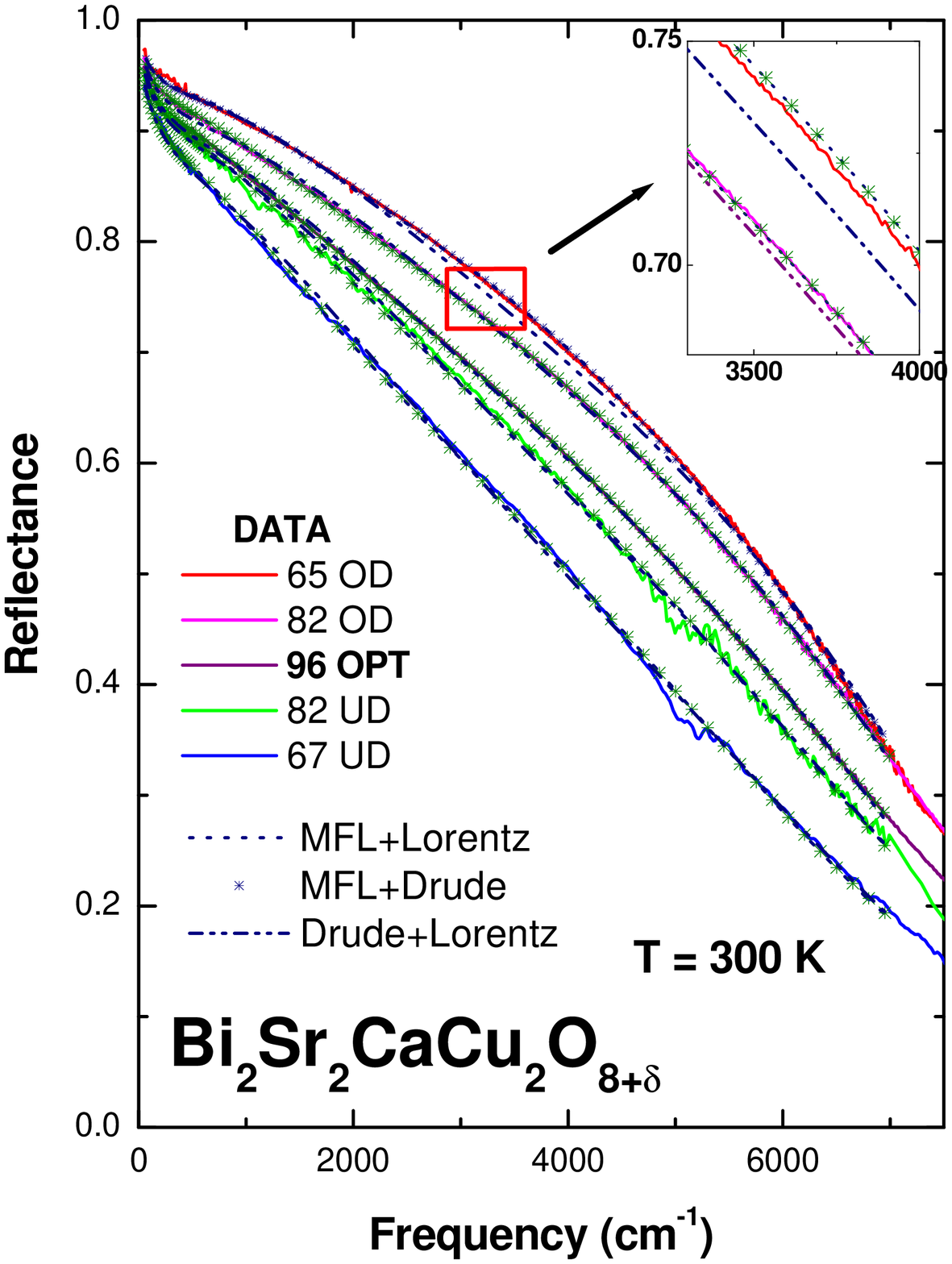}}%
  \vspace*{-1.0cm}%
\caption{Kubo-MFL fits for extended frequency range at 300~K for
Bi$_{2}$Sr$_{2}$CaCu$_{2}$O$_{8+{\delta}}$ at five representative
doping levels: two underdoped ($T_{c}=70$ and 82~K), one optimally
doped ($T_{c}=96$ K), and two overdoped ($T_{c}=82$ and 65~K)
samples. The symbols (MFL+Drude), dotted lines (MFL+Lorentz) and dash-doulbledotted lines (Drude+Lorentz) are the fits and the solid lines are the
corresponding data. The fit parameters are shown in
Table~\ref{tKubo1}. The inset shows a magnified view of the fits and
demonstrates the need for an MFL contribution for a good fit to
the data.}%
\label{Kubograph1}
\end{figure}

%
%
\begin{figure}[t!]
  \vspace*{-0.4cm}%
  \centerline{\includegraphics[width=3.5in]{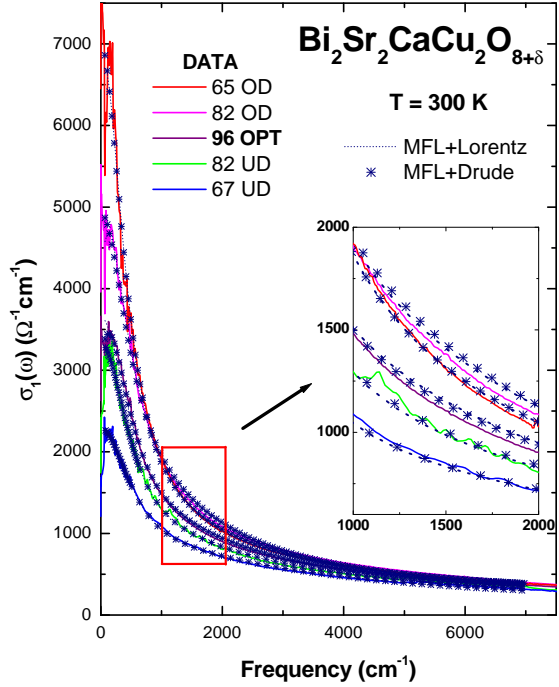}}%
  \vspace*{-1.0cm}%
\caption{Kubo-MFL fits for extended frequency range at 300 K for
Bi$_{2}$Sr$_{2}$CaCu$_{2}$O$_{8+{\delta}}$ optical conductivity at
five representative doping levels: two underdoped ($T_{c}=70$ and
82~K), one optimally doped ($T_{c}=96$ K), and two overdoped
($T_{c}=82$ and 65~K) samples. The symbols (MFL+Drude) and the dotted
lines (MFL+Lorentz) are fits and the solid lines are the corresponding
data. The fit parameters are shown in Table~\ref{tKubo1}.}%
\label{Kubograph2}
\end{figure}

%
%
\begin{figure}[t!]
  \vspace*{-0.4cm}%
  \centerline{\includegraphics[width=3.5in]{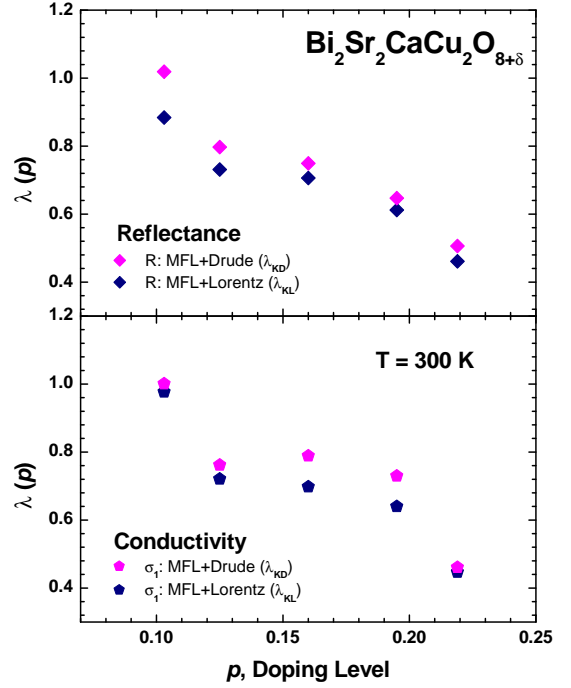}}%
  \vspace*{-1.0cm}%
\caption{Comparison between two fits: MFL channel + a Drude
channel and MFL channel + a Lorentz channel for reflectance and
optical conductivity data.}%
\label{Kubograph3}
\end{figure}

Table~\ref{Bi2212} shows the slope and intercept for each Bi-2212
sample as well as the coupling constant from the $1-R$ slope
analysis between 500 and 1750~cm$^{-1}$ for each doping level, as
well as the coupling constants from Kubo-MFL fits between 60 and
1700~cm$^{-1}$, shown in the $\lambda_{K}(p)$ column.

%
%
\begin{table}
\caption{The table shows the sources of 300~K
Bi$_{2}$Sr$_{2}$CaCu$_{2}$O$_{8+ \delta}$ reflectance $(R)$ data.
We fit the $1-R$ data to a straight line in the frequency range
between 500 and 1750~cm$^{-1}$. The slope and intercept of the
fits are shown in the table. The sample whose $T_{c}$ is 96~K is
in its optimal doping state. It was assigned a doping level of
0.16. We also show the coupling constants from Kubo-MFL fits
between 60 and 1700~cm$^{-1}$ in the table.}

%
%
%

  \begin{ruledtabular}
  \begin{tabular}{ccccccc}
    $T_{c}(K)$ & $p$& slope\ ($\mu$m) & intercept &
    $\lambda_{S}(p)$ & $\lambda_{K}(p)$ & Ref. \\
    \hline
        67&0.103&1.069&0.0818&0.809&0.899& \onlinecite{biu70}\\
        70&0.107&1.232&0.0884&0.927&1.165& \onlinecite{puchkov96b}\\
        82&0.125&0.844&0.0670&0.688&0.720& \onlinecite{puchkov96b}\\
        89&0.144&0.796&0.0620&0.657&0.658& \\
        91&0.160&0.790&0.0630&0.670&0.702& \onlinecite{tu02}\\
        96&0.160&0.803&0.0616&0.681&0.693& new,\onlinecite{eisaki02}\\
        90&0.172&0.669&0.0616&0.585&0.606& \onlinecite{puchkov96b}\\
        82&0.195&0.625&0.0555&0.587&0.606& new,\onlinecite{eisaki02}\\
        73&0.209&0.594&0.0550&0.561&0.584& new \\
        70&0.213&0.552&0.0520&0.524&0.533& \onlinecite{puchkov96b}\\
        65&0.219&0.537&0.0418&0.517&0.457& new \\
        60&0.226&0.564&0.0470&0.526&0.506& new \\
  \end{tabular}
  \end{ruledtabular}
\label{Bi2212}
\end{table}

We were unable to get a fit to the data up to 7000~cm$^{-1}$ with
the simple MFL parameterization of the data. To get a reasonable
fit for an extended range of frequencies we have to add a parallel
Drude channel to the conductivity. We fit the same Bi-2212 data at
300 K from 60 to 7000~cm$^{-1}$ by adding one Drude oscillator.
Here we have three free parameters: the coupling constant
$\lambda$, width and strength of the Drude oscillator $\gamma_{D}$
and $\omega_{pD}$. Fig.~\ref{Kubograph1} shows the data and
Kubo-MFL fits for the extended spectral range and
Table~\ref{tKubo1} shows parameters for the fits. We observe that
the coupling constants [$\lambda_{KD}(p)$] are slightly higher
than those [$\lambda_{K}(p)$] of the previous Kubo-MFL fits
without oscillators. The spectral weight of the Drude oscillator
needed for the good fit shown in the figure ranges from 44 \% of
the MFL spectral weight for the underdoped sample down to 17 \%
for the highly overdoped sample. The monotonic variation of the
added Drude component with doping suggests that the deviations
from the MFL form at high frequency may have a physical
significance.

We also fit the same data between 60 and 7000 cm$^{-1}$ with the
MFL channel and one Lorentz oscillator. Here we have four free
parameters: the coupling constant $\lambda$, center frequency,
width and strength of the Lorentz oscillator $\omega_{L}$,
$\gamma_{L}$ and $\omega_{pL}$. Fig.~\ref{Kubograph1} shows the
data and Kubo-MFL fits for the extended spectral range and
Table~\ref{tKubo2} shows parameters for the fits. The spectral
weight of the Lorentz oscillator needed for the good fit shown in
the figure ranges from 38 \% of the MFL spectral weight for the
underdoped sample down to 13 \% for the highly overdoped sample.
We also calculate the mean square deviation $\chi^{2}$ of the two
high frequency fits to compare the qualities of the fits. The
results are shown in Table~\ref{tKubo3}. With one additional fit
parameter the MFL+Lorentz yields a slight improvement in the fit.
The exact physical meaning of the additional Drude or Lorentz
oscillator is not clear but deviations from the simple MFL
parametrization is expected once the energies begin to approach
the widths of the bands. In Fig.~\ref{Kubograph3} we show the
resulting coupling constants, $\lambda_{KD}(p)$ with
$\lambda_{KL}(p)$ for both reflectance and the optical
conductivity (see the following subsection).

We fit the same reflectance data between 60 and 7000 cm$^{-1}$
with a Drude channel and one Lorentz channel. The fit parameters
and resulting fits are shown in Table~\ref{tKubo21} and
Fig.~\ref{Kubograph1}, respectively. Interestingly in the two
overdoped samples ($T_{c}=82$ K and $T_{c}=65$ K) the least square
fits converge to two Drude oscillators even though we start with
one Drude and one Lorentz oscillator. We also show the mean square
deviations $\chi^{2}_{DL}$ in Table~\ref{tKubo3}. Even though the
Drude+Lorentz has the largest number of parameters it shows the
worst fit. This suggests that an MFL channel is necessary for a
good fit to the Bi-2212 data and probably other cuprates as well.

%
%
\begin{table}
\caption{The MFL+Drude fit parameters of five representative
Bi$_{2}$Sr$_{2}$CaCu$_{2}$O$_{8 + \delta}$ samples: two underdoped
($T_{c}=67$ and 82~K), one optimally doped ($T_{c}$=96~K), and two
overdoped ($T_{c}=82$ and 65~K) samples for both reflectance and
conductivity. The results of the fits are shown in
Fig.~\ref{Kubograph1} and ~\ref{Kubograph2}. We have added to the
MFL conductivity channel (plasma frequency $\omega_{p}$) a Drude
channel with a plasma frequency $\omega_{pD}$ and a width
$\gamma_{D}$. In the table "($R$)" and "($\sigma_{1}$)" stand for
the reflectance and the conductivity, respectively.} %
%
%
%
%
\begin{ruledtabular}
\begin{tabular}{ccccccc}
  $T_{c}$\ (K)& $\omega_{c}$\ (cm$^{-1}$) & $\lambda_{KD}$ &
  $\omega_{p}$\ (cm$^{-1}$) & $\epsilon_{\infty}$ &
  $\gamma_{D}$ & $\omega_{pD}$ \\
  \hline
  67 ($R$) & 7100 & 1.019 & 15040 & 3.14 &  4790 & 9970 \\
  82 ($R$) & 7100 & 0.797 & 16320 & 3.21 &  4250 & 9360 \\
  96 ($R$) & 7100 & 0.749 & 16980 & 3.52 &  4000 & 9960 \\
  82 ($R$) & 7100 & 0.647 & 18789 & 4.16 &  3300 & 10260 \\
  65 ($R$) & 7100 & 0.506 & 19270 & 4.25 &  2930 & 7920 \\ \hline
  67 ($\sigma_{1}$) & 7100 & 1.001 & 15040 & 3.14 &  3980 & 10843 \\
  82 ($\sigma_{1}$) & 7100 & 0.762 & 16320 & 3.21 &  3270 & 10397 \\
  96 ($\sigma_{1}$) & 7100 & 0.789 & 16980 & 3.52 &  2708 & 10987 \\
  82 ($\sigma_{1}$) & 7100 & 0.730 & 18789 & 4.16 &  2013 & 11400 \\
  65 ($\sigma_{1}$) & 7100 & 0.461 & 19270 & 4.25 &  2569 & 9497 \\
\end{tabular}
\end{ruledtabular}

\label{tKubo1}
\end{table}

%
%
\begin{table}
\caption{The MFL+Lorentz fit parameters of five representative
Bi$_{2}$Sr$_{2}$CaCu$_{2}$O$_{8 + \delta}$ samples: two underdoped
($T_{c}=67$ and 82~K), one optimally doped ($T_{c}$=96~K), and two
overdoped ($T_{c}=82$ and 65~K) samples for both reflectance and
conductivity. The results of the fits are shown in
Fig.~\ref{Kubograph1} and ~\ref{Kubograph2}. We have added to the
MFL conductivity channel (plasma frequency $\omega_{p}$) a Lorentz
channel with a center frequency $\omega_L$, a plasma frequency
$\omega_{pL}$ and a width $\gamma_{L}$. In the table "($R$)" and
"($\sigma_{1}$)" stand for the reflectance and the conductivity,
respectively.}
%
%
%
%
%
\begin{ruledtabular}
\begin{tabular}{cccccccc}
  $T_{c}$\ (K)& $\omega_{c}$\ (cm$^{-1}$) & $\lambda_{KL}$ &
  $\omega_{p}$\ (cm$^{-1}$) & $\epsilon_{\infty}$ & $\omega_{L}$ &
  $\gamma_{L}$ & $\omega_{pL}$ \\
  \hline
  67 ($R$) & 7100 & 0.884 & 15040 & 3.14 & 1229 & 5161 & 9344 \\
  82 ($R$) & 7100 & 0.731 & 16320 & 3.21 & 1003 & 4600 & 8910 \\
  96 ($R$) & 7100 & 0.706 & 16980 & 3.52 & 801 & 4177 & 9614 \\
  82 ($R$) & 7100 & 0.612 & 18789 & 4.16 & 741 & 3470 & 9851 \\
  65 ($R$) & 7100 & 0.461 & 19270 & 4.25 & 1472 & 3436 & 6944 \\ \hline
  67 ($\sigma_{1}$) & 7100 & 0.977 & 15040 & 3.14 & 373 & 4115 & 10765 \\
  82 ($\sigma_{1}$) & 7100 & 0.721 & 16320 & 3.21 & 503 & 3585 & 10216 \\
  96 ($\sigma_{1}$) & 7100 & 0.698 & 16980 & 3.52 & 606 & 3007 & 10533 \\
  82 ($\sigma_{1}$) & 7100 & 0.640 & 18789 & 4.16 & 492 & 2322 & 10853 \\
  65 ($\sigma_{1}$) & 7100 & 0.447 & 19270 & 4.25 & 463 & 2722 & 9555 \\
\end{tabular}
\end{ruledtabular}

\label{tKubo2}
\end{table}
%
%
\begin{table}
\caption{The Drude+Lorentz fit parameters of five representative
Bi$_{2}$Sr$_{2}$CaCu$_{2}$O$_{8 + \delta}$ samples: two underdoped
($T_{c}=67$ and 82~K), one optimally doped ($T_{c}$=96~K), and two
overdoped ($T_{c}=82$ and 65~K) samples for reflectance. The
results of the fits are shown in Fig.~\ref{Kubograph1}. We have
five fit parameters: two from a Drude channel with a plasma
frequency $\omega_{pD}$ and a width $\gamma_{D}$ and three from a
Lorentz channel with a center frequency $\omega_L$, a plasma
frequency $\omega_{pL}$ and a width $\gamma_{L}$.}
%
%
%
%
%
\begin{ruledtabular}
\begin{tabular}{cccccccc}
  $T_{c}$\ (K)&$\epsilon_{\infty}$ &$\gamma_{D}$&$\omega_{pD}$ (cm$^{-1}$)&
   $\omega_{L}$ & $\gamma_{L}$ & $\omega_{pL}$ \\
  \hline
  67 ($R$) & 3.14 & 698  & 9149 & 2336 & 8762 & 13200 \\
  82 ($R$) & 3.21 & 618 & 10300 & 2201 & 8493 & 13323 \\
  96 ($R$) & 3.52 & 607 & 10810 & 2022 & 7343 & 13187 \\
  82 ($R$) & 4.16 & 625 & 11958 & 0.14 & 6950 & 14735 \\
  65 ($R$) & 4.25 & 444 & 12268 & 0.12 & 6040 & 14491 \\
\end{tabular}
\end{ruledtabular}

\label{tKubo21}
\end{table}

%
%
\begin{table}
\caption{Comparison between three fits for reflectance data: MFL
channel + a Drude channel, MFL channel + a Lorentz channel and a
Drude channel + a Lorentz channel. We named the mean square
deviations $\chi^{2}_{MD}$ for the MFL+Drude fit, $\chi^{2}_{ML}$
for the MFL+Lorentz fit, and $\chi^{2}_{DL}$ for the
Drude+Lorentz. The additional parameters of the Lorentz fit yields
only a slight improvement when we compare the MFL+Lorentz fit with
the MFL+Drude
fit.} %
%
%
%
%
\begin{ruledtabular}
\begin{tabular}{cccc}
  $T_{c}$\ (K)& $\chi^{2}_{MD}$ & $\chi^{2}_{ML}$& $\chi^{2}_{DL}$ \\
  \hline
  67 & 18.65 & 17.03&21.45  \\
  82 & 16.35 & 15.63&18.62  \\
  96 &  2.21 &  1.48& 6.59  \\
  82 &  3.61 &  2.89& 6.14  \\
  65 &  7.99 &  6.57&43.70  \\
  No. of parameters&3&4&5\\
\end{tabular}
\end{ruledtabular}

\label{tKubo3}
\end{table}

\subsubsection{Conductivity and Kubo-MFL fit}

We also fit the optical conductivity between 60 and 7000 cm$^{-1}$
with an MFL channel and two different oscillators: MFL + Drude or
MFL + Lorentz. We determined the optical conductivity from the
measured reflectance by using a Kramers-Kronig
analysis~\cite{wooten72}, for which extrapolations to $\omega
\rightarrow 0 \:\:\mbox{and} \:\:\infty$ must be supplied. For
$\omega \rightarrow 0$, the reflectance was extrapolated by
assuming a Hagen-Rubens frequency dependence, $(1-R)\propto
\omega^{1/2}$. The reflectance has been extended to high-frequency
by using a literature data~\cite{tarasaki90} and
free-electron behavior ($R\propto \omega^{-4}$). The
result of the fits to the data are shown in Fig.~\ref{Kubograph1}
and the fit parameters are given in Table~\ref{tKubo1} and
~\ref{tKubo2}. We also show the resulting doping dependent
coupling constants from two different fits for both reflectance
and conductivity in Fig.~\ref{Kubograph2}. We observe smoother
doping dependent coupling constant in the result from the reflectance
fit. This is one of the reasons why we use reflectance to study
the doping dependent properties.

The additional Drude and Lorenz oscillators introduced to fit the
wider frequency range data should not be taken literally as
additional physical channels of conductivity. They only serve to
correct the frequency dependence of the MFL spectrum that is known
not to fit well over a wider range of
frequencies~\cite{elazrak94,baraduc96,vanderMarel03}.
%
%
\section{Discussions}
\begin{figure}[t!]
  \vspace*{-0.4cm}%
  \centerline{\includegraphics[width=3.5in]{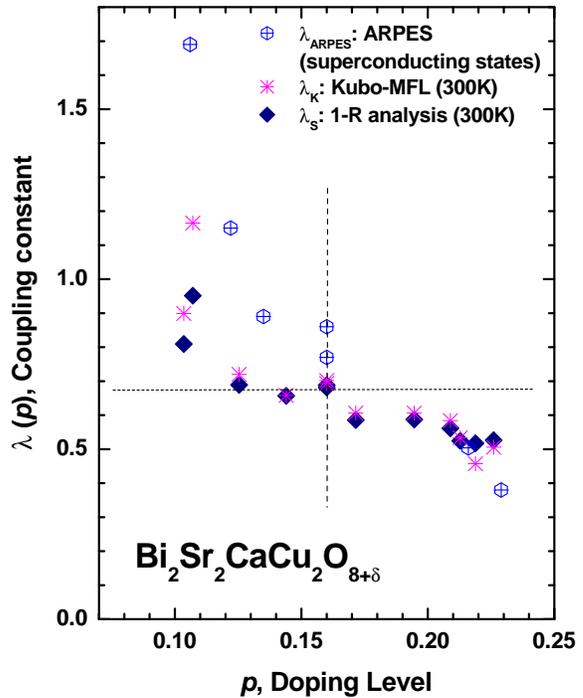}}%
  \vspace*{-1.0cm}%
\caption{Coupling constants, $\lambda (p)$, of
Bi$_{2}$Sr$_{2}$CaCu$_{2}$O$_{8+ \delta}$ at 300~K from two
different methods, discussed in the text and one result from angle
resolved photoemission spectroscopy of Bi-2212 data in the
superconducting state.\cite{johnson01}}%
\label{together}
\end{figure}

We note that as a first approximation, the MFL spectrum of
fluctuations accounts for the data well in the low frequency
region of $\omega < 2000$~cm$^{-1}$ both within the simple $1-R$
slope analysis and the more accurate Kubo formalism. In
Fig.~\ref{together} we compare the dimensionless coupling
constants $\lambda(p)$ from the two different methods of analysis:
1-$R$ and Kubo-MFL fit (without oscillators). The coupling
constant decreases uniformly with doping except for the highly
underdoped samples. However, in the low doping range the pseudogap
temperature $T^*$ may approach our measurement temperature of 300
K. The pseudogap leads to a decreased scattering rate at low
frequencies which would in turn lead to an enhanced slope of the
absorption curve. The upturn in the slope at low doping levels may
therefore well be a pseudogap effect. Thus, our first conclusion
is that at room temperature the coupling constant $\lambda(p)$
decreases uniformly in the doping region that we have studied
without any evidence of any crossovers.

We also show in Fig.~\ref{together} the coupling constants from
the $1-R$ slope analysis and from ARPES.\cite{johnson01} Johnson
{\it et al.}\cite{johnson01} argued that in the normal state the
self-energy is well described by the MFL hypothesis and in the
overdoped region the difference between the superconducting and
normal state dispersion vanishes. They calculated the coupling
constants by using $\lambda = -(\partial \mbox{Re} \Sigma/\partial
\omega)_{E_{F}}$. We have augmented their analysis by estimating
the coupling constant at room temperature which is lower than
their published low-temperature value in the underdoped region.
This is probably due to a rearrangement of spectral weight of the
spin-fluctuation spectrum with temperature.  As the temperature is
lowered, this rearrangement takes two forms: spectral weight is
removed from low frequencies as a result of the development of the
spin gap but added to intermediate frequencies with the
development of the neutron resonance that couples strongly to the
carriers. The net result is a strong increase in $\lambda$ which
suggests that spectral weight is removed from high frequencies to
fill the neutron mode.\cite{schachinger03}

The overall good agreement of the coupling constant and its doping
dependence between the infrared and the ARPES data is surprising
since the infrared response represents an average over the Fermi
surface, whereas the ARPES data shown in Fig.~\ref{together}
represents corrections to the self energy in the $(\pi,\pi)$
direction.  Also, ARPES measures the particle life-times directly,
whereas infrared is weighted in favor of large angle scattering.

%
%

Our data on the intercept and the curvature does show evidence of
doping-dependent discontinuities.  The effect of the pseudogap on
the intercept would be opposite from that on the slope, causing it
to be smaller at low doping if there was a pseudogap. Instead, as
shown in Fig.~\ref{intercept}, we observe the opposite effect in
the underdoped region, the intercept is higher than what would be
expected from a uniformly increasing $\lambda$ with underdoping.
Also, in the highly overdoped region the intercept drops, whereas
the slope appears to have a more uniform trend in the 0.16 to 0.22
doping region. The overall result of these effects is to produce
an S-shaped curve centered at optimal doping.

However, the clearest evidence of a singular doping level at
$p=0.16$ comes from our analysis of the doping dependence of the
curvature plotted in Fig.~\ref{curvature}.  Here we see a good fit
to a straight line that goes through zero exactly at optimal
doping $p=0.16$.  This phenomenon is directly related to the well
known observation that the dc resistivity in high temperature
superconductors varies linearly with temperature only at optimal
doping and curved on either side of optimal doping. This was shown
by Takagi {\it et al.}\cite{takagi92} for overdoped and underdoped
La$_{2-x}$Sr$_x$CuO$_4$, by Carrington {\it et al.} in underdoped
YBa$_2$Cu$_3$O$_{7-\delta}$ (Ref.~\onlinecite{carrington93}), by
Kubo {\it et al.} in overdoped Tl$_2$Ba$_2$CuO$_2$
(Ref.~\onlinecite{kubo91}) and by Konstantinovic {\it et al.} for
underdoped Bi-2212\cite{konstantinovic99} and for underdoped and
overdoped
Bi$_{2}$Sr$_{1.6}$La$_{0.4}$CuO$_{y}$\cite{konstantinovic01}. The
present work extends this work to Bi-2212 and covers both
overdoped and underdoped materials for the same system.  Needless
to say, we study the {\it frequency dependence} whereas the dc
resistivity work is based on the temperature dependence. Thus we
show by an independent method that at room temperature the MFL
fluctuations are centered at optimal doping $p=0.16$.

%
%

\section{Conclusions}
We have analyzed the 300~K reflectance data of Bi-2212 single
crystals at various doping levels from underdoped ($p=0.103$) to
highly overdoped ($p=0.226$). We found three smoothly varying
quantities with doping, the doping dependent slope, intercept and
the curvature of the reflectance in the relaxation region of
frequencies. From the smooth trend of the reflectance slope with
doping we derive a useful measure of the doping level.  We
estimate a reliable doping dependent dimensionless coupling
constant $\lambda_{S}(p)$ from the $1-R$ slope analysis which
varies smoothly from 0.93 to 0.52 in the doping range from
$p=0.103$ to 0.226. We also obtained doping dependent curvature of
$1-R$. Plotting the curvature as a function of doping we found
that the curvature goes through zero at optimal doping.

%
%

\acknowledgments

We would like to acknowledge useful discussions with E. Abrahams,
J.P.~Carbotte, M.R.~Norman and C.M.~Varma. This work has been
supported by the Canadian Natural Science and Engineering Research
Council and the Canadian Institute of Advanced Research. Work at
Brookhaven was supported by U.S. Department of Energy under
Contract No. DE-AC02-98CH10886.

%
%

\end{document}